\documentclass[twocolumn,prb,showpacs,preprintnumbers,amsmath,amssymb,superscriptaddress]{revtex4}% PRB

\usepackage{graphicx}% Include figure files

\usepackage[active]{srcltx}
\def\tr{\mathrm{tr}}
\newcommand{\e}{\varepsilon}
\newcommand{\dbar}{\kern-.1em{\raise.8ex\hbox{ -}}\kern-.6em{d}}

\def\lin{L}
\def\lind{{\cal L}}

\newtheorem{thm}{Theorem}%[section]

\def \be{\begin{equation}}
\def \ee{\end{equation}}
\def \bea{\begin{eqnarray}}
\def \eea{\end{eqnarray}}

\usepackage[usenames]{color}
\begin{document}
\title{Optimal parametrizations of adiabatic paths}
%%%%%%%%%%%%%%%%%%%%
\author{J.~E.~Avron}
%\\
\affiliation{Department of Physics, Technion, 32000 Haifa, Israel}
\author{M. Fraas}
\affiliation{Department of Physics, Technion, 32000 Haifa, Israel}
\author{G.M. Graf}
\affiliation{Theoretische Physik, ETH Zurich, 8093 Zurich, Switzerland}
\author{P. Grech}
\affiliation{Theoretische Physik, ETH Zurich, 8093 Zurich, Switzerland}
%%%%%%%%%%%%%%%%%%%%%%%
\date{\today}%
\begin{abstract}

We show that, given a path of Hamiltonians and a fixed time to complete it, there are many parametrizations, i.e. time-tables for running along the path, such that the ground state of the initial Hamiltonian is mapped {\em exactly} on  the final ground state. In contrast, we show that if dephasing  is added to the dynamics then there is a {\em unique} parametrization which maximizes the fidelity of the final state with the target ground state. The optimizing parametrization solves a variational problem of Lagrangian type 
and has constant tunneling rate along the path irrespective of  the gap. Application to quantum search algorithms recovers the
Grover result for appropriate scaling of the dephasing with the size of the data base. Lindbladians that  describe wide open systems require special care  since they may mask hidden resources that enable beating the Grover bound.
 \end{abstract}
\maketitle

In the theory of adiabatic quantum control \cite{shore} and
quantum computation \cite{farhi}, one is interested in reaching a
target state from a (different) initial state with high fidelity,
as quickly as possible, subject to given cap on the available
energy. %This leads to an optimization problem which we shall now
%describe more precisely.
%This involve \cite{cerf, vazirani, seiler} path optimization.
The initial state is assumed to be the ground state of a given
Hamiltonian $H_0$ and the target state is the ground state of a
known Hamiltonian $H_1$. The two are connected by a smooth
interpolating path in the space
of %(self-adjoint)
Hamiltonians. The interpolation is denoted by $H_q$ with
$q\in[0,1]$. An example is the linear interpolation
    \be\label{eq:path}
    H_q= (1-q)H_0+q H_1, \quad (0\le q\le 1).
    \ee
%More complicated interpolations are sometimes useful
However, any smooth interpolation \cite{farhi2} 
%which guarantees the boundedness of the
%energy resources  and depends smoothly on $q\in(0,1)$ 
will do.  We are interested in the optimal parametrization of the interpolating path.  
That is, a time-table for the path which optimizes the fidelity of the time-evolving state at the end time with the ground state of the target Hamiltonian.
We will show that when the evolution is unitary, there is no unique optimizer--there are plenty of them. However, when dephasing is added to the dynamics there is a unique optimizer which we characterize and discuss.   

For the sake of simplicity we assume that the
Hilbert space has a dimension $N$ (finite) and that $H_q$ is a
self-adjoint matrix-valued function of $q$ with ordered  simple eigenvalues $e_a(q)$, so that
    	\be\label{eq:Gamma1}
	H_q =\sum_{a=0}^{N-1} e_a(q) P_a(q)   \,.
	\ee
$P_a(q)$ are the corresponding spectral projections.

A slow change of $q$ tends to maintain the system in its ground
state up to an error due to tunneling. We are interested in
getting as close as possible to the target state within the time
$\cal{T}$ allotted  to traversing the path. The controls at our
disposal are a. The total time $\cal{T}$ and b. The
parametrization of the path $q(s)=q_\e(s), \ s\in[0,1]$ for given $\e$.  Here $s=\e t$ is
the \emph{slow time} parametrization and $\e=1/\cal{T}$ the
adiabaticity parameter.

The cost function is the tunneling $ T_{q,\e}(1)$ at the end point, where $ T_{q,\e}(s)$ is defined by
 \be\label{eq:tunneling}
    T_{q,\e}(s)=1-\tr\bigl(P_0(q)\rho_{q,\e}(s)\bigr).
    \ee
$\rho_{q,\e}(s)$ is the quantum state at slow-time $s$ which has
evolved from the initial condition $\rho_{q,\e}(0)=P_0(0)$.

A related but different  optimization problem commonly considered
in quantum information is to optimize {\em upper bounds} on the
tunneling  \cite{seiler}. The difference is that the cost function is evaluated not for a fixed, given interpolation, but for the worst case for {\em any} (smooth) interpolation between {\em any} two
Hamiltonians belonging to certain classes.

We consider two types of evolutions: (a) Unitary
evolutions generated by $H_q$. (b) Non-unitary evolutions
generated by appropriate Lindblad generators $\lin_q$ \cite{Davies}. Since (a)
is a special case of (b),  both  are of the form
    \be\label{eq:L}
    \e \dot \rho =\lin_q(\rho),
    \ee
where $\dot{}=d/ds$ and
    \be\label{eq:lindblad}
    \lin(\rho)=-i[H,\rho]+ \sum_{j=1}^M \left(2\Gamma_j\rho\Gamma^*_j- \Gamma_j^*\Gamma_j\rho -
    \rho\Gamma_j^*\Gamma_j\right)
    \ee
with %$H$ self-adjoint and
$\Gamma_j$, a-priori, arbitrary.
Adiabatic evolutions are a singular limit of the evolution
equations since $\e$ hits the leading derivative. Unitary
evolutions are generated when $\Gamma_j=0$.

In the case of unitary evolution the optimization problem has no
unique solution, on the contrary, optimizers are ubiquitous. More
precisely:
\begin{thm}\label{thm:nonuniqueness}
Let
    \be\label{eq:H}
        2 H_q = \mathbf{g}(q)\cdot\sigma
    \ee
be any smooth interpolation of a 2-level system where $\sigma$ is
the vector of Pauli matrices and $ \mathbf{g}(q)$ a smooth,
vector valued function with a gap, $ |\mathbf{g}(q)|\ge g_0>0$; let 
$\e/g_0$ be small. Then, in
a neighborhood of order $\varepsilon$ of any smooth parametrization,
there are many
non-smooth parametrizations with zero tunneling and therefore
many smooth parametrizations with arbitrarily small tunneling.
\end{thm}

\begin{figure}[htb]
%\centering
\hskip -1 cm 
\includegraphics[height=6 cm]{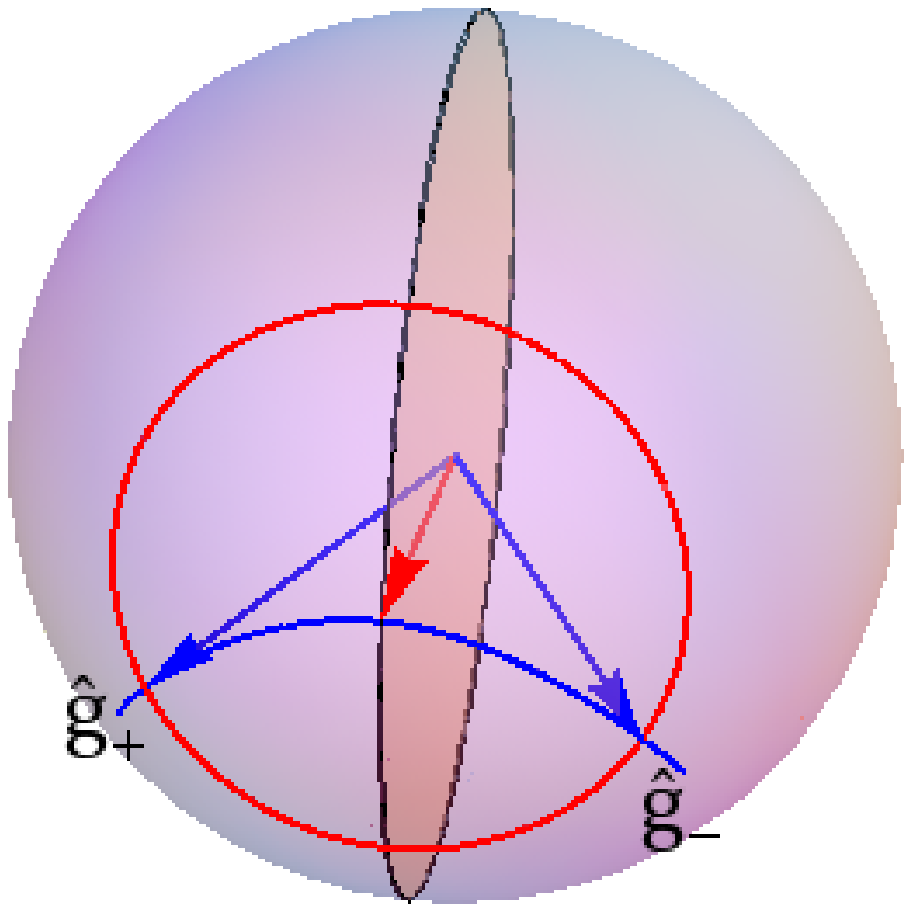} \hskip 0 cm
\includegraphics[height=5cm]{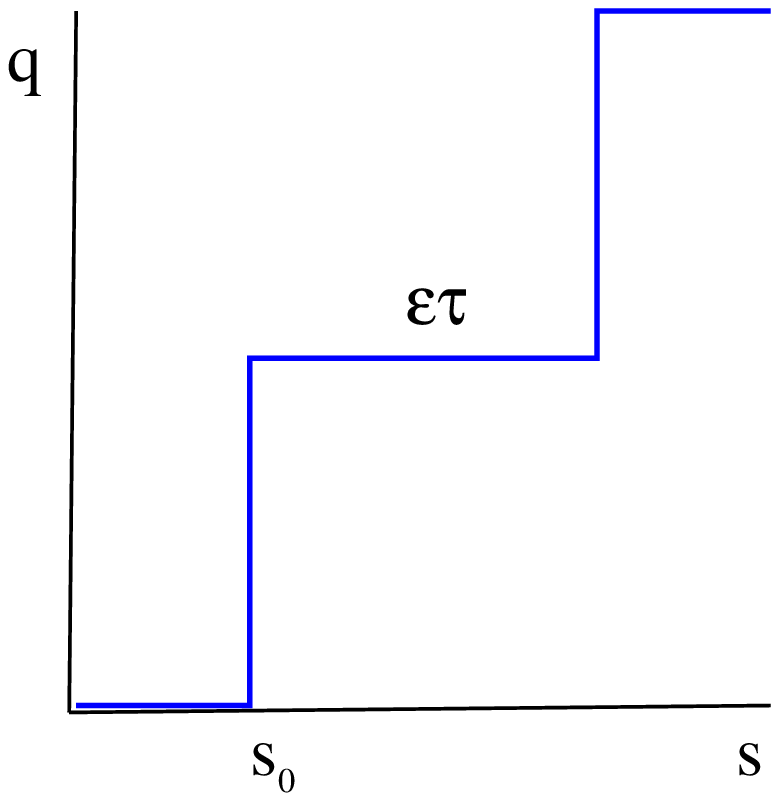}
\caption{Left: $\hat{\mathbf{g}}_\pm$ are the images on the Bloch
sphere of the end points of an interval of size $O(\e )$ of  a
given parametrization (blue). The intersection of the associated
interpolating path with the equatorial plane (shaded) determines
the point $q^*$ and thereby the axis of precession
$\hat{\mathbf{g}}(q^*)$  (red) that maps the instantaneous state at
the initial end point to the corresponding state at the final end
point. Right: A non-smooth interpolating path that takes the
instantaneous eigenstate at the beginning of the interval to the
instantaneous eigenstate at the end of the interval with no
tunneling.}\label{fig:bloch}
\end{figure}

We shall sketch the main idea behind the proof. Consider a
discretization of any given parametrization to (slow) time
intervals of size $2\pi \e/g_0$. In each interval one can find a
point $q^*$, such that the time-independent Hamiltonian $H_{q^*}$
acting for appropriate time $\tau\le 2\pi/|\mathbf{g}(q^*)|\le 2\pi/g_0$,
will map the image on the Bloch sphere of the starting point
$q_-$ to the image of the end point $q_+$.
This says that there are many (non-smooth) paths, labelled by the
continuous parameter $s_0$ in Fig.~\ref{fig:bloch}, that map the
instantaneous state at the initial end point to the corresponding
state at the final end point. These paths have zero tunneling. The
existence of $q^*$ follows from the geometric construction in
Fig.~\ref{fig:bloch}: $\mathbf{g}(q^*)$ is a point of intersection
of the path  with the
equatorial plane orthogonal to
$\hat{\mathbf{g}} (q_+)-\hat{\mathbf{g}}(q_-)$. The resulting parametrization
differs from the original one by at most
$(\sup_s|\dot q(s)|)\cdot 2\pi\e/g_0$, as
seen from the mean-value theorem.

Dephasing Lindblad operators belong to a special class of Lindblad operators which share with unitary evolutions the existence of $N$ stationary states. (In contrast with generic Lindblad operators that have a unique equilibrium state.) More
precisely,   $\lind$ is a  dephasing Lindblad operator, if all the
spectral projections $P_a$ of $H$ are stationary states, namely
$P_a\in \ker\lind$.
This is the case when $[\Gamma_j,H]=0$, and the condition
is also necessary  when $H$ has simple eigenvalues, as can
be seen by expanding $\tr(P_a\lind(P_a))=0$.
We can then write
   % \be\label{eq:Gamma2}
$\Gamma_j=\sum_{a=0}^{N-1} \sqrt{\gamma}_{j a} P_a,$
%\quad  j=1,\dots , N^2-1
  %  \ee
where $\sqrt{\gamma}$ is a rectangular, $M\times N$, matrix
(without loss, $M=N^2-1$). It follows that dephasing
Lindbladians have the form \cite{alicki}:
    \be\label{eq:dephasing}
    \lind(\rho)=-i[H,\rho]+
    \sum_{a,b} 2 \gamma_{ba}\,P_a\rho P_b-
    \sum_a\gamma_{aa}\big\{P_a,\rho\big\} ,
    \ee
where  $0\le\gamma$ is  a positive matrix.  Time-dependent dephasing Lindblad operators \cite{aks}
are then defined by setting $H\to H_q$ and $P_a\to P_a(q)$ and $\gamma\to \gamma(q)$ .

The motion of $ \ker\lind_q$ with $q$ can be interpreted
geometrically as follows: The space of (unnormalized) states is a
fixed $N^2$ dimensional convex cone.  The {\em normalized} instantaneous stationary states are
a simplex whose vertices are the instantaneous spectral
projections $P_a(q)$.  This simplex
rotates with $q$ like a rigid body, since the vertices remain
orthonormal, $\tr(P_aP_b)=\delta_{ab}$ and the motion is
purely orthogonal to the kernel,  $\tr(P'_a P_b)=0$ where
$P'_a=dP_a/dq$. This follows from the fact that for orthogonal projections
 $P'_a$ is off-diagonal
     \be\label{eq:dotP}
     P'_a(q)=\sum_{b \neq c} P_b(q) P'_a(q) P_c(q).
         \ee

An adiabatic theorem for dephasing Lindblad operators can be inferred from \cite{nenciu}. It  says that
%%%%%%%%%%%%%%%%%%%%%%%%%%%%%%%%%%%%%%
%	 \begin{thm}
%
 %Let $\lind_q$ be a smooth family of dephasing Lindblad operators with
 %(smooth) Hamiltonian $H_q$. Let $P_a(q)$ be the instantaneous spectral
%projections for the simple eigenvalues of $H_q$.
 %Then 
 the solution $\rho_{q,\e}^{(a)}$ of the adiabatic
 evolution, Eq.~(\ref{eq:L}), for the parametrization $q(s)$ and initial condition
 $\rho_{q,\e}^{(a)}(0) = P_a(0)$, adheres to
 the instantaneous spectral projection\footnote{Since there are several
 energy scales in the problem: $\e$, $\gamma$ and the minimal $g_0$,
  the remainder term is guaranteed to be small provided
 $\e\ll \gamma, g_0$ is the smallest energy  scale.}
    \be
    \rho_{q,\e}^{(a)}(s) =P_a(s) +O(\e),\quad (s>0).
    \label{eq:afgg}
    \ee
    %\end{thm}

For the sake of writing simple formulas we shall, from now on,
restrict ourselves to the special case where the positive matrix
$\gamma(q)>0$ of Eq.~(\ref{eq:dephasing}) is a multiple of the identity
    \be\label{eq:DL}
    \lind_q(\rho)
    =-i[H_q,\rho]-\gamma(q) \sum_{j \neq k} P_j(q) \rho
    P_k(q).
    %\label{eq:31}
    \ee

Our main results follow from:
%%%%%%%%%%%%%%%%%%%%%%%%%%%%%%%%%%%%%
	\begin{thm}
 	\label{thm:afgg}
Let $\lind_q$ be the dephasing Lindblad of Eq.~(\ref{eq:DL}),
and $\rho_{q,\e}$ a solution of (\ref{eq:L}) with initial
condition $\rho(0) = P_0(0)$ for the parametrization $q(s)$.
Assume a gap condition
 $e_a(q)\neq  e_b(q) ,\,(a \neq b)$.
Then the tunneling defined by Eq.~(\ref{eq:tunneling}), is given
by
    \be
    T_{q,\e}(1) = 2 \e  \int_0^1 {M(q)} \, \dot q^2 \,  ds + O(\e^2),\quad
    \label{eq:trf}
    \ee
where the $q$ dependent \emph{mass} term
    \be\label{eq:mass}
 {M(q)}=\sum_{a \neq 0} \,
\frac{\gamma(q)\,\tr(P_a {P'_0}^2 )} {(e_0(q) - e_a(q))^2 +
\gamma^2(q)} \ge 0
    \ee
is independent of the parametrization.  $P'_0(q)$ denotes a
derivative with respect to $q$ and $\dot q(s)$  with respect to $s$.
\end{thm}
In  the special case of a 2-level system, Eq.~(\ref{eq:H}), where
$\mathbf{g}(q)$ is a 3-vector valued function parametrized by its
length
%    \be
    $ d\mathbf{g}(q)\cdot
    d\mathbf{g}(q)=(dq)^2 $
 %  \ee
the ``mass'' term of Eq.~(\ref{eq:mass}) takes the simple
form
    \be\label{eq:massTwoLevels}
    M(q) =
    \frac{\gamma(q)} 4  \, \frac{|\hat{
    \mathbf{g}}'|^2(q)\,}{{g}^2(q) + \gamma^2(q)}
    \ee
$|\hat {\mathbf{g}}'|$ is the velocity w.r.t. $q$ on the Bloch
sphere ball and $g(q)=|\mathbf{g}(q)|$ is the gap.

%%%%%%%%%%%%%%%%%%%%%%%%%%%%%%%%%%%%%%%%%

\emph{Remark: }  For a 2-level system undergoing \emph{unitary} evolution a similar variational principle to Eq.~(\ref{eq:trf}), but with a different  $M(q)$,  was proposed, as an ansatz,  in \cite{lidar} for the purpose of determining an optimal
path, rather than an optimal parametrization of a given path.

Before proving the theorem let us discuss some of its consequences:
%\item
Note first, that the tunneling rate, $ {2 \e {M(q)}\dot q^2}\ge 0$, is local and uni-directional. It follows that whatever has tunneled can not be recovered, in contrast with unitary evolutions.
%  \item
Eq.~(\ref{eq:trf})  has the standard form of variational
Euler-Lagrange problems with a Lagrangian that is proportional to the adiabaticity $\e$ and with the interpretation of kinetic energy with position dependent mass. This variational problem has   a unique minimizer $q_0(s)$ in the adiabatic limit,
%    \item
in contrast with the case for unitary evolutions, which by Theorem \ref{thm:nonuniqueness} has no unique minimizer.
%  \item

Since the Lagrangian is $s$ independent $q_0(s)$ conserves
``energy''  and the tunneling rate is constant along the minimizing
orbit. This gives a local algorithm for optimizing the parametrization: Adjust the speed $\dot q (s)$ to keep the tunneling
rate constant.
%    \item
The optimal speed along the path is then
    \be\label{eq:OptimalSpeed}
    \dot  q=\sqrt{\frac{\tau}{M(q)}},
    \ee
where $\tau>0$ is a normalization constant. This formula
quantifies the intuition that the optimal velocity is large when
the  gap is large and the projection on the instantaneous ground
state changes slowly.
  %\item
The optimal tunneling, $T_\mathrm{min}$,  is then
    \be\label{eq:OptimalTunneling}
    T_\mathrm{min}=2\e \tau+O(\e^2),\quad
   \sqrt{\tau}=\int_0^1\ {dq}\,{\sqrt{M( q)}}.
   \ee
This formulas will play a role in our analysis of Grover search algorithm.
%\end{itemize}
%\end{enumerate}

%%%%%%%%%%%%%%%%%%%%%%%%%%%%%%
We now turn to proving Theorem \ref{thm:afgg}. Evidently
    \be
    1 - \tr(P_0 \rho_{q,\e})(1) = -\int_0^1 \frac{d}{ds}
\tr\big(P_0(q) \rho_{q,\e}(s)\big) \,d s\,.
    \ee
Using Eq.~(\ref{eq:L}),  the defining property of dephasing Lindbladians, $\lind_q(P_0(q))=0$, and by Eq.~(\ref{eq:dephasing}), the concomitant  $\lind_q^*(P_0(q))=0$, one finds
    \be\label{eq:integrand}
  \frac{d}{ds}
\tr\big(P_0(q) \rho_{q,\e}(s)\big)=\tr\bigl({P'_0}(q)\, \rho_{q,\e}(s)\bigr)\ \dot q(s)\,.
 \ee
Now, the identity,
    \be
    \lind^*(P_a A P_b) =
\big(i(e_a - e_b) -\gamma\big) P_a A P_b,\quad (a \neq b)
    \ee
together with Eq.~(\ref{eq:dotP})  shows  that
    \be X = \sum_{a \neq b}
\frac{P_a P'_0 P_b}{i(e_a - e_b) - \gamma}. \label{eq:x}
    \ee
solves the equation
    \be
    {P}_0'(q) =\lind_q^*\big(X(q)\big) \label{eq:druha}
    \ee
Substituting this in Eq.~(\ref{eq:integrand}) gives the  identity
 \be
  \frac{d}{ds}
\tr\big(P_0(q) \rho_{q,\e}(s)\big)=
\e\, \tr\big(X(q)\,\dot{\rho}_{q,\e}(s)\big)\,\dot q(s).
 \label{eq:tra}
 \ee
Integrating by parts the last identity gives an expression
involving $\rho$ but no $\dot \rho$. This allows us to use the adiabatic theorem
and replace $\rho$ by $P +O(\e)$. We then undo
the integration by parts to get Theorem \ref{thm:afgg}.
%%%%%%%%%%%%%%%%%%%%%%%%%%%%%%%%%%%%%

In the theory of Lindblad operators  $H$ and $\Gamma_j$ of Eq.~(\ref{eq:lindblad}) can be chosen independently.  However, as we shall now show, if one makes some natural assumptions about the bath, the dephasing rate $\gamma$ of Eq.~(\ref{eq:DL}) is constrained by the gaps of $H$.

To see this we turn to quantum search with dephasing \cite{aks,boixo}.
Grover has shown \cite{grover} that $O(\sqrt N)$ queries of an
oracle suffice to search an unstructured data base of size $N\gg
1$. The adiabatic formulation of the problem leads to the study of
a 2-level system with a small gap given by
\cite{seiler,cerf}
    \be\label{eq:gap}
    g^2(q)= 4 \frac{(1-q)q}N+\left(1-2q\right)^2
    \ee
and large velocity on the Bloch sphere
    \be\label{eq:asympt}
    |\hat {\mathbf{g}}'(q)|= \sqrt{\frac 1 N-\frac{1}{N^2}}\ \frac{2}{ g^2(q)} .
   \ee

The time scale $\tau$, which determines the optimal tunneling, can
be estimated by evaluating the integrand in
Eq.~(\ref{eq:OptimalTunneling}) at its maximum, $q=1/2$, and
taking the width  to be $1/\sqrt N$. This gives
    \be\label{eq:tau}
    \tau =O\left(\frac{M(1/2)}N\right)
    \ee
to leading order in the adiabatic approximation.

The adiabatic formulation \cite{farhi} fixes the scaling of the
minimal gap $g_0\sim \frac 1 {\sqrt N}$ but does not fix the
scaling of the dephasing rate $\gamma$ with $N$.
 We shall now
address the issue of what physical principles determines the
scaling of the dephasing with $N$. To this end we consider various cases.

%\begin{itemize}
 %   \item
The regime $\gamma \ll \e$ is outside the
framework of the adiabatic theory described here, but is close to
the unitary scenario, \cite{farhi,seiler}. For the adiabatic expansion
and Eq.~(\ref{eq:tau}) to hold  $\e \ll \gamma$. This means that 
in case of small dephasing, $\gamma\ll g_0$,
the allotted time, ${\cal T} \gg \gamma^{-1}\gg O(\sqrt N)$, is longer
than Grover search time.  For such times the theory developed here can be
used to estimate the tunneling, but it is not appropriate for
optimizing the search time. To optimize the search time one
needs to study bounds on the tunneling rather than a first order
term in $\e$.
%This is similar to what one does the
%approach of unitary adiabatic searches .

   % \item
When  dephasing is comparable to the gap, $\gamma\sim g_0$, one finds $M(1/2)\sim 1/g^3_0$ and from
    Eqs.~(\ref{eq:tau}, \ref{eq:gap}) one recovers Grover's result for the search time
        \be
        {\cal T} =O\left(\frac 1 {g^3_0 N}\right)=O(\sqrt N).
        \ee
   % \item
Finally, consider the dominant dephasing case: $\gamma\gg g_0$. Here $M\sim \gamma^{-1}/g^2_0$ and from Eqs.~(\ref{eq:tau}, \ref{eq:gap}) one
    finds
        \be\label{eq:surprise}
        {\cal T}=O\left(\gamma^{-1}\right)\,.
        \ee
If $\gamma$ scaled like $\gamma\sim N^{-\alpha/2},\ 1
> \alpha $, then ${\cal T}=O\left( N^{\alpha/2}\right)$ which seems to beat
Grover time. %This intriguing result requires a proper interpretation.

 %\end{itemize}

The accelerated search enabled by strong dephasing
is in apparent conflict with the optimality of Grover bound  \cite{fg,bennett}: Consider the
Hamiltonian dynamics of the joint system and bath, which underlies
the Lindblad evolution. By an argument of \cite{cerf} for a
universal bath, the Grover search time is optimal. How can one reconcile
Eq.~(\ref{eq:surprise}) with this result?
Before doing so, however, we want to point out that
Eq.~(\ref{eq:surprise}) is not an
artefact of perturbation theory: While $T_\mathrm{min}=2\e\tau$ is
valid in first order in $\e$, an estimate $T_\mathrm{min}
\lesssim\e\tau$, with $\tau$ as in
Eq.~(\ref{eq:OptimalTunneling}), remains true for all $\e$
provided $\gamma\gtrsim g_0$.

The resolution is that a Markovian
bath with $\gamma \gg g_0$ can not be universal and must be system
specific: The bath has a premonition of what the solution to the
problem is. (Formally, this ``knowledge'' is reflected in the dephasing in the instantaneous eigenstates of $H_q$.)  Lindbladians with  dephasing rates that dominate the gaps
mask resources hidden in the bath.  This can also be seen by the following argument: Dephasing
can be interpreted as the monitoring of the observable $H_q$. The
{\em time-energy} uncertainty principle \cite{popescu} says that
if $H_q$ is unknown, then the rate of monitoring is
bounded by the gap. The accelerated search occurs when monitoring rate
exceeds this bound, which is only possible if the bath already
``knows'' what $H_q$ is.  When $H_q$ is known, the bath can freeze the system in the
instantaneous ground state arbitrarily fast.  Consequently,  the Zeno effect  
\cite{zeno} then allows for the speedup of the evolution without paying a large price in tunneling.

The formal theory of Lindblad operators allows one to choose the operators,  $H$ and $\Gamma_j$ in  Eq.~(\ref{eq:lindblad}), independently. From the discussion in the last paragraph one learns that one must exercise care in using Lindbladians for systems that are wide open.
%, i.e. where $H$ is small and $\Gamma_j$ are large.
Markovian baths which are universal, i.e. oblivious of the state of the system, give rise to dephasing Lindbladians,  with dephasing rates that are bounded by the spectral gaps of the system.

{\bf Acknowledgments.} This work is supported by the ISF and the
fund for Promotion of research at the Technion. M.~F. was
partially supported by UNESCO fund.  
We thank J. \AA berg, S. Popescu and N.~Yoran
for useful discussions, N. Lindner for bringing to our
attentions ref.~\cite{boixo} and thank A.~Elgart and
D.~Krej\v{c}i\v{r}\'{i}k for the observation that unitary evolutions
have no good minimizers.


\begin{thebibliography}{99}
\bibitem{shore} K.~Bergmann, H.~Theuer, B.W.~Shore,
%Coherent population transfer among quantum states of atoms and molecules, 
Rev. Mod. Phys. \textbf{70}, 1003-1025 (1998);
%\bibitem{rabitz}
R.S. Judson and H. Rabitz, 
%Teaching lasers to control molecules,
Phys. Rev. Lett. \textbf{68}, 1500-1503 (1992).

%\bibitem{Davies} E.B.~Davies,
%Quantum Stochastic Processes II, Commun. math. Phys. 19, 83.105 (1970)

\bibitem{farhi} E.~Farhi, J.~Goldstone, S.~Gutmann, M.~Sipser,
%Quantum computation by adiabatic evolution, 
quant-ph/0001106 (2000);
%\bibitem{aharonov}
D.~Aharonov, W.~van~Dam, J.~Kempe, Z.~Landau, S.~Lloyd,
O.~Regev, 
%Adiabatic quantum computation is equivalent to standard quantum computation, 
FOCS \textbf{45}, pp.42-51 (2004),
quant-ph/0405098

%\bibitem{averin} DV Averin, Adiabatic quantum computation with cooper pairs,
%Solid State Communications, 1998
\bibitem{farhi2} E.~Farhi, J.~Goldstone, S.~Gutmann,
%Quantum adiabatic evolution algorithms with different paths,
quant-ph/0208135.
\bibitem{seiler} S.~Jansen, M.B.~Ruskai, R.~Seiler,
%Bounds for the adiabatic approximation with applications to quantum computation, 
J.  Math.  Phys. \textbf{48}, 102111 (2007);
A.~Ambainis, M.B.~Ruskai, Report of workshop: Mathematical
aspects of quantum adiabatic approximation.

\bibitem{Davies} E.B.~Davies, Quantum Theory of Open Systems, Academic Press, (1976);  
E.B.~Davies, 
%Markovian Master Equations, 
Commun. Math.
Phys. {\bf 39}, 91-110 (1974).
\bibitem{alicki} R. Alicki and M. Fannes, Quantum dynamical systems,
Oxford (2001).

\bibitem{aks} J. \AA berg, D. Kult, E. Sj\"oqvist, 
%Robustness of the adiabatic quantum search, 
Phys. Rev. A {\bf 71}, 060312(R) (2005); quant-ph/0412124.


\bibitem{nenciu} G. Nenciu, G. Rasche, 
%On the adiabatic theorem for nonself-adjoint Hamiltonians. 
J. Phys. A {\bf 25}, 5741-5751 (1992); W.K.~Abou~Salem, 
%On the quasi-static evolution of nonequilibrium steady states, 
Ann. H. Poincar\'{e} \textbf{8}, 569-596 (2007); A. Joye, 
%General adiabatic evolution with a gap condition, 
Comm. Math. Phys.
{\bf 275}, 139-162 (2007).


\bibitem{lidar}
A.T.~Rezakhani, W.J.~Kuo, A.~Hamma, D.A.~Lidar, P.~Zanardi,
%Quantum adiabatic brachistochrone, 
Phys. Rev. Lett. \textbf{103},
080502 (2009)

%\bibitem{afgg1} J.E.~Avron, M. Fraas, G.M. Graf, P. Grech,
%Landau-Zener tunneling for dephasing Lindblad evolutions, arXiv:0912.4640.

\bibitem{boixo} S.~Boixo, E.~Knill, R.~Somma, 
%Eigenpath traversal by phase randomization, 
Quantum Inf. and Com. \textbf{9}, 833-855 (2009); arxiv:0903.1652

 \bibitem{grover} L.K. Grover, %Quantum Mechanics helps in searching for a needle in a haystack,
Phys. Rev. Lett. \textbf{79}, 325-328 (1997); quant-ph/9706033.

\bibitem{cerf} W.~van~Dam, M.~Mosca, V.~Vazirani, 
%How powerful is adiabatic quantum  computation?, 
Proc. 42nd FOCS, 279-287 (2001); quant-ph/0206003.; J.~Roland, N.J.~Cerf, 
%Quantum search by local adiabatic evolution, 
Phys. Rev. A \textbf{65}, 042308 (2002).
%\bibitem{vazirani}

\bibitem{popescu}Y. Aharonov, S. Massar, S. Popescu, 
%Measuring energy, estimating Hamiltonians, and the time-energy uncertainty relation,
Phys. Rev. A \textbf{66}, 052107 (2002).

\bibitem{zeno} W.M.~Itano, D.J.~Heinzen, J.J.~Bollinger, D.J.~Wineland, 
%Quantum zeno effect, 
Phys. Rev. A
 \textbf{41}, 2295-2300 (1990).



\bibitem{fg} E.~Farhi, S.~Gutmann,
%An analog analogue of digital quantum computation,
Phys. Rev. A {\bf 57}, 2403 (1998); quant-ph/9612026.

\bibitem{bennett} C.~H.~Bennett, E.~Bernstein, G.~Brassard, U.~Vazirani,
%Strengths and weaknesses of quantum computing, 
quant-ph/9701001.

 \end{thebibliography}
\end{document}